# EXERGAMES FOR TELEREHABILITATION

## THESIS

Submitted in Partial Fulfillment of

the Requirements for

the Degree of

**MASTER OF SCIENCE** (Mechatronics and Robotics)

at the

NEW YORK UNIVERSITY

TANDON SCHOOL OF ENGINEERING

by

**Satish Reddy Bethi**

**May 2020**

# EXERGAMES FOR TELEREHABILITATION

**THESIS**

Submitted in Partial Fulfillment of

the Requirements for

the Degree of

**MASTER OF SCIENCE** (Mechatronics and Robotics)

at the

NEW YORK UNIVERSITY

TANDON SCHOOL OF ENGINEERING

by

**Satish Reddy Bethi**

**May 2020**

Approved:

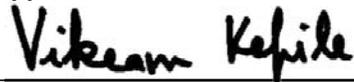

Prof. Vikram Kapila, Advisor

May 20, 2020

Date

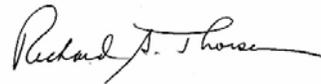

Prof. Richard Thorsen, Dept. Chair

May 20, 2020

Date

University ID: N10296494
Net ID:      srb629

# VITA

Satish Reddy Bethi was born in Hyderabad, India, on 17$^{th}$ of August 1997. He received his bachelor's degree in Mechanical Engineering from Jawaharlal Nehru Technological University, Hyderabad (JNTUH). In the Fall of 2018, he began working on his master's thesis at the Mechatronics, Controls, and Robotics Lab, Department of Mechanical and Aerospace Engineering, New York University, Tandon School of Engineering. Throughout his graduate studies, he has worked on developing exergames and simulations for medical devices and blockchain technology, funded by the NYU Tandon School of Engineering Graduate Student Assistantship/Fellowship and the TREAT grant NIH P2CHD086841.

# ACKNOWLEDGEMENT

I express my gratitude and appreciation to Prof. Vikram Kapila and Dr. Preeti Raghavan for providing opportunity, guidance, and funding throughout my graduate studies. I would like to thank the PhD students Ashwin RajKumar, Hassam Khan Wazir, and Sai Prashant Krishnamoorthy for helping me through every step of my journey. Finally, I would like to acknowledge my co-authors Fabio Vulpi, Veena Jayasree-Krishnan, Sahil Kumar, Renu Karthick Rajaguru Jayanthi, and Dr. Shramana Ghosh for their contributions, which supplemented my research skills to perfection.

# ABSTRACT

**EXERGAMES FOR TELEREHABILITATION**

By

**Satish Reddy Bethi**

**Advisor: Prof. Vikram Kapila**

Submitted in Partial Fulfillment of the Requirement for

The Degree of Master of Science (Mechatronics and Robotics)

May 2020

Recent advancements in technology have improved the connectivity between humans enhancing the transfer of information. Leveraging these technological marvels in the healthcare industry has led to the development of telehealth allowing patients and clinicians to receive and administer treatment remotely. Telerehabilitation is a subset of telehealth that facilitates remote rehabilitation treatment for patients. Providing rehabilitative services to the aging baby boomer population requires tech-savvy solutions to augment the therapists and clinicians for effective remote monitoring and tele-medicine. Hence, this thesis develops easy-to-use exergames for low-cost mechatronic devices targeting rehabilitation of post-stroke patients. Specifically, it demonstrates wearable inertial sensors for exergames consisting of an animated virtual coach for providing patients with instructions for performing range of motion exercises. Next, a gaming environment is developed for task-specific rehabilitation such as eating. Finally, exergames are developed for rehabilitation of pincer grasping. In addition to gamified interfaces providing an engaging rehabilitation experience to the user, the data acquired from the mechatronic devices facilitate data-driven telerehabilitation.

# TABLE OF CONTENTS

# List of Figures

# List of Tables

# Chapter 1 Introduction

The recent trends in miniaturization and open-source embedded systems have improved the accessibility to low-cost microcontrollers (e.g., Arduino). These system-on-chip controllers are integrated with special communication modules (e.g., WiFi, Bluetooth, and cellular) yielding a small footprint microcontroller that is widely applicable for usage in medical and wearable technology applications. Use of low-cost sensors and wearable devices can enhance health monitoring and enable improved personalized treatment of patients by the clinicians.

Recent advancements in micro-electromechanical systems have led to the miniaturization of inertial measurement units (IMU) and magnetometer, accelerometer, and gyroscope (MARG) sensor arrays, collectively referred to as inertial sensors. These inertial sensors are widely used in diverse applications (e.g., aerospace, automotive, sports medicine, etc.) for pose measurement and have revolutionized the ability to precisely track the position and orientation of a rigid body in 3D space. Moreover, the small footprint of such sensors permits their integration in everyday ubiquitous devices such as smartphones. With recent innovations in wireless communication protocols obviating the need for wired connectivity and allowing portability, new avenues for adoption of inertial sensors have opened in myriad wearable electronic devices such as smart watches. These inertial sensors, coupled with wireless connectivity, are also used for motion capture (MOCAP) applications [1]–[3].

## 1.1 Challenges with traditional stroke rehabilitation

Most tasks constituting activities of daily living (ADL) have minimum range of motion (ROM) requirements at various joints [4]. Restoring the ability to perform ADLs in individuals with impaired movement therefore requires clinicians to assess ROM and customize the exercises to each patients' activity limitations. Commercially available devices such as goniometers [5], inclinometers [6], and video-graphic methods [7] are used by therapists to assess patient's ROM in one-on-one clinical settings. Goniometers and inclinometers have limited inter-observer agreement due to variability in positioning the sensors on the patient's body and can capture motion only for one joint at a time. Video-graphic methods also show low inter-observer agreement due to differences in camera positions and often require extensive post-acquisition data analysis. Furthermore, these methods are not suitable for remote assessments and individualized treatments, which are essential to enhance accessibility.

MOCAP is an interdisciplinary research topic that focuses on quantifying motion and enabling interaction in real and virtual environments. Commercially available MOCAP systems can be broadly classified into: (*i*) optical marker-based systems [8], (*ii*) electromagnetic position tracking system [9], (*iii*) markerless optical systems [10], [11], and (*iv*) inertial sensing systems [1]–[3]. Marker-based optical MOCAP is the gold standard for tracking joint position and angular movement with high precision and accuracy [8]. However, such systems require precise marker placement and expensive cameras, all of which are burdensome for clinical use. Furthermore, marker occlusion can occur during limb movements, making tracking difficult. Electromagnetic position tracking (e.g., by Ascension Technology Corp.) computes the position of body-worn electromagnetic sensors relative to a transmitter [9]. These systems avoid the use of multiple cameras and marker occlusion, but they are not easy to use for clinical purposes. Markerless optical systems include devices such as the Kinect$^{TM}$ V2 (Microsoft Corp.,

Redmond, WA), which is a popular MOCAP device to measure joint positions in 3D space [10], [11]. However, data from the Kinect and other markerless video-analysis systems cannot make measurements in the horizontal plane, such as shoulder internal-external rotation and forearm pronation-supination, which are critical for ADL [4]. Furthermore, the Kinect cannot be used in noisy visual environments. Recent advancements in deep learning with markerless MOCAP using videography can reduce the human effort to track human and animal behavior [12], [13], but these have the same limitations as other vision based systems such as the Kinect, and do not provide precise triplanar measurements for real-time applications.

## 1.2 Application of Exergames

The use of rehabilitative devices and telerehabilitation reduce the cost for rehabilitation. However, patient motivation is critical for efficient recovery [14]. The repetitive and tiresome tasks required for rehabilitation must be performed regularly and may discourage the patient leading to discontinuity in rehabilitation therapy. Incorporating an element of entertainment through gamification of these activities can stimulate the patient to perform the activities while playing the game. These types of games that include physical activity are called "exergames" [15]. The exergames add additional visual feedback to the patients enabling them to perceive their incorrect movements or mistakes. Such exergames can aid the therapists by allowing them to measure the patient performance using the score and set difficulties based on it. Moreover, recent availability of free game engines, such as Unity and Unreal, and their integration with commercially available technologies (e.g., Kinect) as in [16], has shown promise in home-based telerehabilitation.

This work aims to develop exergames for *in-situ* mechatronics-based low-cost rehabilitative devices to provide home-based telerehabilitation. Chapter 2 discusses the exergame and its interfaces for wearable inertial sensors (WIS) for upper extremity motion capture and range of motion assessment. Chapter 3 shows the exergame environment and its interfaces for performing functional sub-tasks of eating using a RehabFork system. In Chapter 4, we discuss the exergame and its interfaces for a grasp rehabilitation device for performing lifting and grasping tasks. Chapter 5 summarizes the contributions of this thesis with future research suggestions.

# Chapter 2 WISE: Wearable Inertial Sensors for Exergames

## 2.1 Introduction

Arm movements are crucial for performing several ADL. Each ADL has minimum ROM requirements for the upper extremity (UE) joints to successfully complete the task [4]. However, neurological events such as stroke, multiple sclerosis, spinal cord injury, nerve damage, etc., can limit an individual's ROM, which in turn prevents them from performing several ADLs and lowers their quality of life. The pathway for recovery of lost motor skills includes: (*i*) determining the movement limitations to facilitate development of a treatment plan and (*ii*) gauging recovery to tailor treatment changes based on patient progress. The anatomy of the human arm with seven degrees of freedom requires advanced motion capture systems to measure the complex movements at the shoulder, elbow, and wrist.

Existing tools for ROM assessment used in clinical practice include: (*i*) hand-held measurement devices such as a goniometer [17], inclinometer [6], etc.; and (*ii*) video analysis software such as the Dartfish [18]. However, these devices do not capture all the degrees of freedom of arm motion. In a research setting, several commercially available motion capture devices are used that can be broadly classified under: (*i*) marker-based optical motion capture systems; (*ii*) electromagnetic position tracking systems; (*iii*) markerless motion capture systems; and (*iv*) inertial sensing systems. However, these systems require trained personnel for setup and data processing, which prohibits their translation from research to clinical practice and home-based environments.

A review of wearable sensors for applications to rehabilitation is provided in [19], which outlines the importance of deploying wearable technologies in home and clinical environments for data-driven rehabilitation. Moreover, providing healthcare rehabilitative services for the aging baby boomer population requires tech-savvy solutions to augment the therapists and clinicians for effective remote monitoring and telemedicine. Video and computer gaming facilitate an entertaining and engaging user experience while performing monotonous repetitive exercises and improve the therapeutic benefits of the treatment [20]. This chapter reviews the use of wearable inertial sensors (WIS) combined with an exergame to visualize the triplanar limb movements and facilitate home-based rehabilitation. Several subsections in Section 2.1 are adopted from [21] and included here for completeness.

**Wearable Inertial Sensors**

The availability of open-source game engines (such as Unity and Unreal Engines) have provided a reliable platform for 3D visualization and improved the accessibility for development of gamification software. Furthermore, companies such as Adobe Mixamo provide high quality 3D models for use in these gaming environments. This, combined with low cost mechatronic-based devices using microcontrollers such as Arduino, has introduced new possibilities for home-based rehabilitation using exergames.

**MARG Sensors**

The WIS system uses five wearable sensor modules each consisting of a MARG sensor fitted in a 3D printed enclosure. Four UE arm sensors (left, right, upper, and lower arm segments) and one back-mounted sensor are used. The WIS modules, each fitted with a BNO055 inertial sensor [22], require an initial calibration of their built-in MARG sensors for accurate absolute orientation measurement. Furthermore, precise mounting on the human body is crucial for accurate measurement of ROM. These sensor modules are integrated with Bluetooth-enabled microcontrollers called RFduino to stream their absolute orientation quaternions relative to the earth's magnetic and gravitation fields. The transmitted data stream is received by a receiver that sends the data to the game through a serial port.

The angles are usually represented as Euler angles (including Tait-Bryan angles), quaternions, and axis/angle representations. Although, Euler angles are the easiest to understand, they often lead to gimbal lock and singularity problems. Moreover, the use of quaternions by Unity for object rotations and the computational simplicity of quaternions spurred us to use quaternions for rotation description.

In brief, a unit quaternion represents the rotation of a coordinate frame in 3D space by an angle $\theta$ about a unit vector $V = [V_x \quad V_y \quad V_z]^T$. Equation (2.1) shows the representation of a unit quaternion.

$$q_V(\theta) := \left(\cos\frac{\theta}{2} \quad V_x \sin\frac{\theta}{2} \quad V_y \sin\frac{\theta}{2} \quad V_z \sin\frac{\theta}{2}\right) \tag{2.1}$$

A unit quaternion $q := (q_w \quad q_x \quad q_y \quad q_z)$ includes a scalar part $q_w$ and a vector part $Q := [q_x \quad q_y \quad q_z]$. Throughout this chapter, uppercase alphabets such as Q denote vectors and lowercase alphabets such as q denote quaternions. Consistent with the notation in prior literature [2], [3], we use "$\otimes$" to denote quaternion product defined by equation (2.2) and "$*$" to denote conjugate defined by equation (2.3).

$$p \otimes q := (p_0 q_0 - P \cdot Q \quad p_0 Q + q_0 P + P \times Q) \tag{2.2}$$

$$q^* := (q_0 \quad -Q) \tag{2.3}$$

Next, equation (2.4) represents the inverse of a quaternion.

$$q^{-1} := \frac{1}{\|q\|} q^* \Rightarrow q \otimes q^{-1} = q^{-1} \otimes q = (1 \quad 0 \quad 0 \quad 0) \tag{2.4}$$

The results of a pre-clinical study on testing the usability and accuracy of the WIS system in contrast to an alternative motion capture technology is presented in [16]. Figure 2-1 shows a user wearing the WIS system in neutral pose and the calibration holder for the WIS system. Our prior experimental study leveraged MATLAB-based user-interfaces (UIs) for data acquisition, calibration, and *in-situ* sensor mounting on the user's body [21]. The MATLAB-based UIs were best suited for experimentation with researchers, but less suitable for use by patients and therapists. Moreover, another prior study found that the use of an animated virtual coach for ROM training improves system usability [16]. Our design motivation for the exergame is to develop a holistic application that integrates multiple interfaces for sensor calibration, sensor mounting, and sensor data collection to assist patients and clinicians. Furthermore, such data-driven rehabilitative approaches will enable clinicians to tailor personalized exercises to patients' needs paving a pathway for precision rehabilitative treatment.

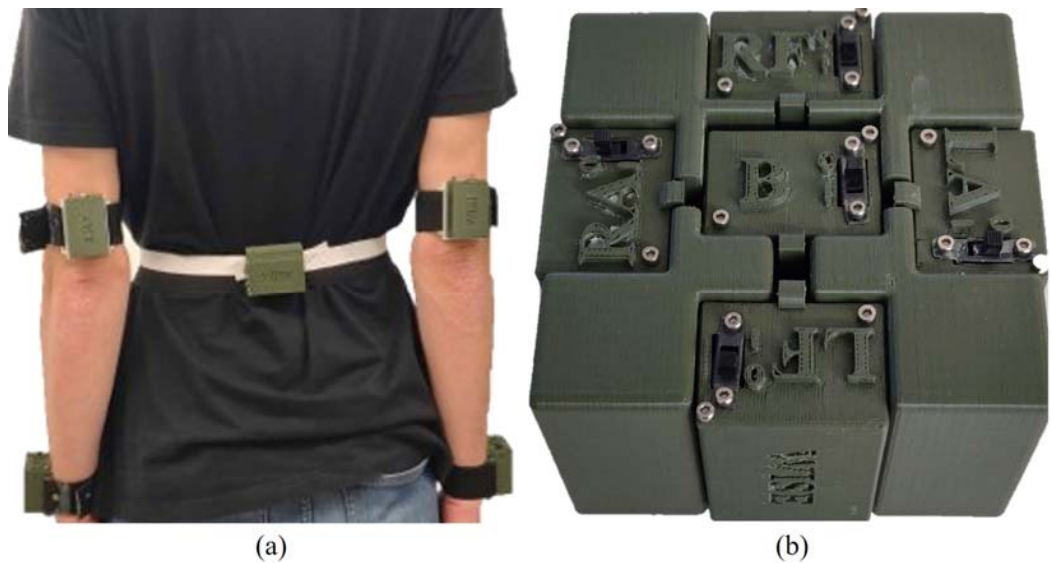

Figure 2-1: (a) WIS modules worn by a user for range of motion assessment and (b) WIS modules placed in a sensor calibration holder

**Quaternions transformations to fit an avatar's input**

Unity supports 3D models with rigs and allows animation of these models using the underlying rigs. Rigging is a process of attaching a bone structure to a 3D mesh to enable animation of the 3D model. Rigging is commonly used in computer graphics for animating 3D objects in video games and movies. The rig available for our avatar can be seen in Figure 2-2. However, for our purpose, we use only the arms, forearms, and two spine joints for animation of the avatar and we link it with the arm, forearm, and the back sensors. However, the default coordinate system in Unity is left-handed and the model starts with a T-pose. Figure 2-3 shows the coordinate frames of the devices and the avatar in a T pose. Thus, to use the data transmitted from the WIS, we perform quaternion transformation from the WIS frame to the avatar frame.

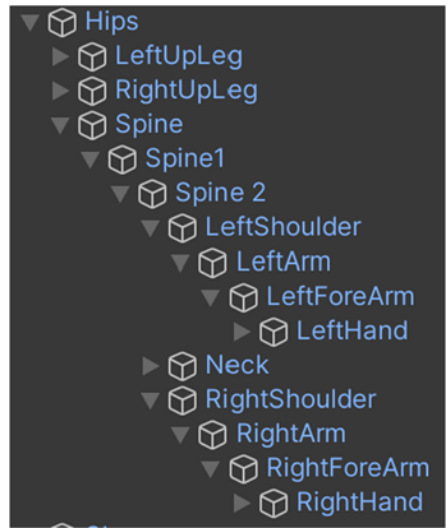

Figure 2-2: Bones available for the avatar

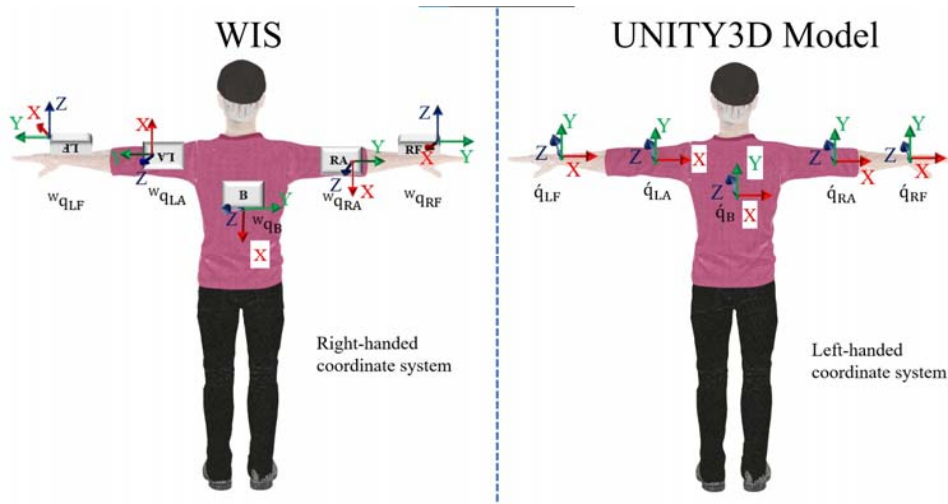

Figure 2-3: Co-ordinate frames of the devices and the avatar in a T pose

We assume that using [21], $\hat{q}_B, \hat{q}_{LA}, \hat{q}_{RA}, \hat{q}_{LF}$, and $\hat{q}_{RF}$ are produced as corrected outputs of sensor quaternions ${}^W q_B$, ${}^W q_{LA}$, ${}^W q_{RA}$, ${}^W q_{LF}$, and ${}^W q_{RF}$ (from Figure 2-3) for the back, left arm, right arm, left forearm, and right forearm, respectively, where ${}^W q_{(\cdot)}$ denotes the quaternion of the IMU $(\cdot)$ expressed in the world frame W (i.e., $\mathcal{F}_W$). Note that for a given unit vector $V = [V_x \ V_y \ V_z]^T$, $q(V) = (0 \ V^T)$ denotes the quaternion corresponding to V and V(q) represents the vector part of the quaternion q. The left arm quaternion relative to the back (i.e., $\tilde{q}_{LA}$) is obtained as below, where we first rotate $\hat{q}_B$ by 180° about the Z axis of the back to produce $\hat{q}_{LATrans}$, which is the quaternion that maps $\tilde{q}_{LA}$ to $\hat{q}_{LA}$ as shown in the following equations.

$$\hat{Z}_W = [0 \ \ 0 \ \ 1] \tag{2.7}$$

$$\hat{q}_{\hat{Z}_B} = \hat{q}_B \otimes q(\hat{Z}_W) \otimes \hat{q}_B^* \tag{2.8}$$

$$\hat{q}_{LATrans} = \hat{q}_{\hat{Z}_B}(\pi) \otimes \hat{q}_B \tag{2.9}$$

$$\tilde{q}_{LA} = \hat{q}_{LATrans}^{-1} \otimes \hat{q}_{LA} \tag{2.10}$$

Note that $\hat{q}_{\hat{Z}_B}(\pi)$ in (2.9) is computed using the expression given in (2.1). Similarly, $\tilde{q}_{RA}$ the right arm quaternion relative to the back is obtained from equation (2.11).

$$\tilde{q}_{RA} = \hat{q}_B^{-1} \otimes \hat{q}_{RA} \tag{2.11}$$

The left forearm quaternion relative to the left arm (i.e., $\tilde{q}_{LF}$) is obtained by rotating $\hat{q}_{LA}$ by 90° about the Y axis of the left arm to produce the transformation $\hat{q}_{LFTrans}$, which is the quaternion that maps $\tilde{q}_{LF}$ to $\hat{q}_{LF}$ as shown below.

$$\hat{Y}_W = [0 \quad 1 \quad 0] \tag{2.12}$$

$$\hat{q}_{\hat{Y}_{LA}} = \hat{q}_{LA} \otimes q(\hat{Y}_W) \otimes \hat{q}_{LA}^* \tag{2.13}$$

$$\hat{q}_{LFTrans} = \hat{q}_{\hat{Y}_{LA}}\left(\frac{\pi}{2}\right) \otimes \hat{q}_{LA} \tag{2.14}$$

$$\tilde{q}_{LF} = \hat{q}_{LFTrans}^{-1} \otimes \hat{q}_{LF} \tag{2.15}$$

Similarly, $\tilde{q}_{RF}$ is quaternion for the right forearm relative to the left arm. It is obtained by rotating $\hat{q}_{RA}$ by $-90°$ about the Y axis of the right arm to produce the transformation $\hat{q}_{RFTrans}$, which is the quaternion that maps $\tilde{q}_{RF}$ to $\hat{q}_{RF}$ as shown below.

$$\hat{q}_{\hat{Y}_{RA}} = \hat{q}_{RA} \otimes q(\hat{Y}_w) \otimes \hat{q}_{RA}^* \tag{2.16}$$

$$\hat{q}_{RFTrans} = \hat{q}_{\hat{Y}_{RA}}\left(-\frac{\pi}{2}\right) \otimes \hat{q}_{RA} \tag{2.17}$$

$$\tilde{q}_{RF} = \hat{q}_{RFTrans}^{-1} \otimes \hat{q}_{RF} \tag{2.18}$$

For the back, $\tilde{q}_B$ is the quaternion used in unity model. It is obtained by rotating $\hat{q}_B$ by 90° about the Y axis of the back to produce the transformation $\hat{q}_{BTrans}$, which is the quaternion that maps $\tilde{q}_B$ to $\hat{q}_B$.

$$\hat{q}_{\hat{Y}_B} = \hat{q}_B \otimes q(\hat{Y}_w) \otimes \hat{q}_B^* \tag{2.19}$$

$$\hat{q}_{BTrans} = \hat{q}_{\hat{Y}_B}\left(\frac{\pi}{2}\right) \otimes \hat{q}_B \tag{2.20}$$

$$\tilde{q}_B = \hat{q}_{BTrans}^{-1} \otimes \hat{q}_B \qquad (2.21)$$

Next, we convert the transformed quaternions $\tilde{q}_{LA}$, $\tilde{q}_{RA}$, $\tilde{q}_{LF}$, $\tilde{q}_{RF}$, and $\tilde{q}_B$ from the right-handed coordinate system to the left-handed coordinate system. This change of coordinate system is performed for each joint quaternion by extracting its corresponding angle $\theta_{\{Joint\}}$ and unit vector $V_{\{Joint\}} = [V_{\{Joint\}_x} \; V_{\{Joint\}_y} \; V_{\{Joint\}_z}]^T$ where {Joint} represents each joint. Then, the unit vector $V_{\{Joint\}}$ of each quaternion is remapped as

$$V'_{LA} = [-V_{LA_y} \; V_{LA_x} \; -V_{LA_z}]^T \qquad (2.22)$$

$$V'_{RA} = [V_{RA_y} \; -V_{RA_x} \; -V_{RA_z}]^T \qquad (2.23)$$

$$V'_{LF} = [-V_{LF_y} \; V_{LF_z} \; V_{LF_x}]^T \qquad (2.24)$$

$$V'_{RF} = [V_{RF_y} \; V_{RF_z} \; -V_{RF_x}]^T \qquad (2.25)$$

$$V'_B = [V_{B_y} \; -V_{B_x} \; -V_{B_z}]^T \qquad (2.26)$$

Finally, the new quaternions $\acute{q}_{\{Joint\}}$ for each joint in the left-handed coordinate system of Unity are obtained by using $\acute{q}_{\{Joint\}} = q_{V'_{\{Joint\}}}(-\theta_{\{Joint\}})$.

**Joint coordinate system**

The quaternions are an effective representation for rotation and computation in 3D space, however, they are rarely used to characterize ROM measurements by therapists and clinicians. The JCS is a standard reporting method proposed by the International Society of Biomechanics (ISB) for computing human joint angles [23], [24]. Furthermore, reporting results using a single standard allows transparent communication between researchers and clinicians. The JCS method uses the proximal coordinate frame as a reference to define the joint angle of the distal coordinate frame. We adopt the method proposed in [23] for computing the joint angles of the UE. The shoulder joint angles use the thorax coordinate frame as the reference and the elbow joint angles use the shoulder coordinate frame as the reference. The coordinate frames for all the sensor modules in a starting neutral pose ($\mathbb{N}_P$) are shown in Figure 2-4.

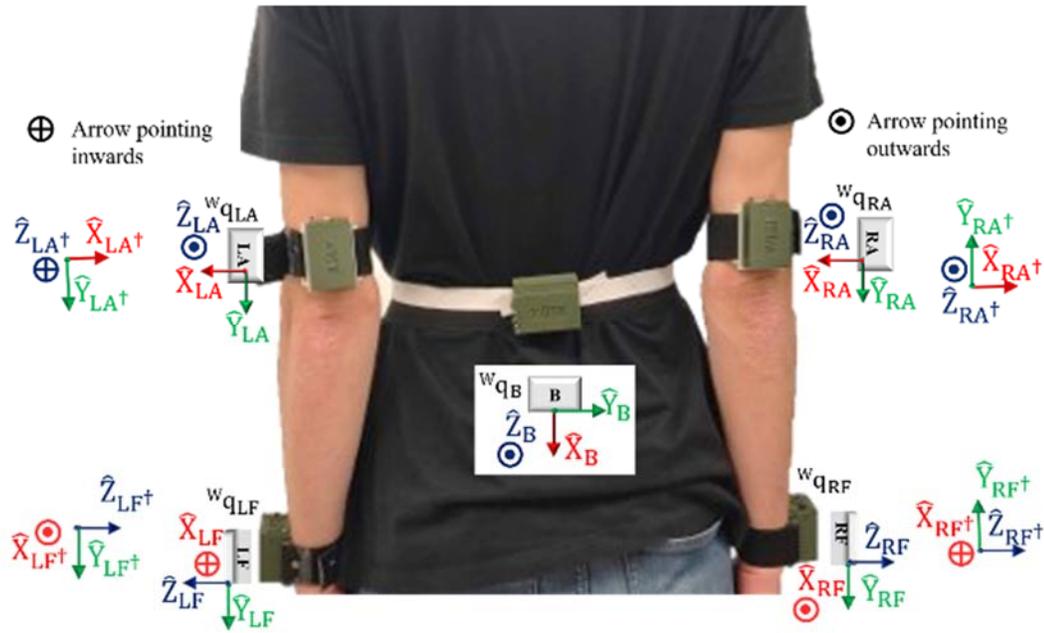

Figure 2-4: Coordinate frames of all sensor modules in neutral pose ($\mathbb{N}_P$)

In the JCS implementation of the WIS system, the back-sensor module B is used as a reference for LA and RA sensor modules to compute the shoulder joint angles. Similarly, the LA and RA sensor modules are used as references for the LF and RF sensor modules, respectively, to compute the elbow and forearm movements. For the shoulder angle computation, an initial reference is needed for the back inertial sensor module at $\mathbb{N}_P$. To do so, two quaternions $q_{RBref}$ and $q_{LBref}$ are created as shown below in (2.27).

$$q_{RBref} = q_{LBref} = \hat{q}_{\hat{Z}_B}\left(-\frac{\pi}{2}\right) \otimes \hat{q}_B \tag{2.27}$$

The sign convention of shoulder joint angle measurements is defined as extension (−) and flexion (+), adduction (−) and abduction (+), and external (−) and internal (+) rotation. The axes shown in Figure 2-4 for the $\hat{q}_{LA}$, $\hat{q}_{RA}$, $q_{LBref}$, and $q_{RBref}$ are rotated by 180° to achieve a similar sign convention. All the rotated coordinate frames are pictorially represented in Figure 2-4. The quaternion representing the rotation of the shoulder relative to the back WIS module is extracted by using (2.29) and (2.28) for the left and right sides, where $(\cdot)^\dagger$ denotes the quaternions for the aforementioned rotated coordinate frames.

$$q_{LS} = q^*_{LBref\dagger} \otimes q_{LA\dagger} \tag{2.28}$$

$$q_{RS} = q^*_{RBref\dagger} \otimes q_{RA\dagger} \tag{2.29}$$

The $Y - X - Y'$ Euler angle convention is used in [23] to obtain the shoulder joint angles. Since the orientation of the LA and RA WIS modules differ from [23], $Y - Z - Y'$ Euler angle convention is adopted. The joint angles are computed using MATLAB's built-in command quat2angle from $q_{LS}$ and $q_{RS}$. The quat2angle command returns angles $\theta_Y$, $\theta_Z$, and $\theta_Y + \theta_{Y'}$ that represent rotation in the shoulder plane, shoulder elevation, and shoulder internal-external rotation, respectively. Shoulder elevation $\theta_Z$ refers to shoulder flexion-extension (in the sagittal plane) when $\theta_Y \approx 90°$ and to shoulder abduction-adduction (i.e., the frontal plane) when $\theta_Y \approx 0°$.

The JCS implementation for measuring elbow rotation requires the use of left arm (LA) and right arm (RA) inertial sensors as references, i.e., $q_{LAref}$ and $q_{RAref}$, respectively, which are computed as below.

$$q_{LAref} = \hat{q}_{\hat{Y}_{LA}}\left(\frac{\pi}{2}\right) \otimes \hat{q}_{LA} \tag{2.30}$$

$$q_{RAref} = \hat{q}_{\hat{Y}_{RA}}\left(-\frac{\pi}{2}\right) \otimes \hat{q}_{RA} \tag{2.31}$$

The sign convention for the elbow and forearm measurements are defined as extension $(-)$ and flexion $(+)$ and supination $(-)$ and pronation $(+)$. As per above, the axes shown in Figure 2-4 for the coordinate frames $q_{RAref}$, $q_{LAref}$, $q_{LF}$, and $q_{RF}$ are rotated by 180° to achieve a similar sign convention. The relative quaternions representing the left ($q_{LE}$) and right ($q_{RE}$) elbow joint angles are computed as below.

$$q_{LE} = q^*_{LAref\dagger} \otimes q_{LF\dagger} \tag{2.32}$$

$$q_{RE} = q^*_{RAref\dagger} \otimes q_{RF\dagger} \tag{2.33}$$

Next, as in [23], the Z-X-Y Euler angle convention is used to obtain the left and right elbow joint angles by using quat2angle MATLAB command from $q_{LE}$ and $q_{RE}$, respectively. The quat2angle command returns angles $\theta_Z$, $\theta_X$, and $\theta_Y$ that indicate elbow flexion-extension, carrying, and pronation-supination angles, respectively. The carrying angle is the angle between the humerus in the upper arm and the ulna in the forearm, which ranges between 8° to 20° [25], [26].

**WIS mounting and alignment**

Mounting the sensors at the distal end of the limb segment reduces most errors in measurement. For example, the forearm sensors (LF and RF) are placed proximal to the wrist joint to produce acceptable results for elbow rotation. However, even when the arm sensors (LA and RA) are placed just proximal to the elbow joint, they are prone to erroneous measurements of internal-external rotation at the shoulder due to skin movements. Thus, correct mounting of the WIS modules is critical for accurate measurement of joint ROM. Inertial sensors have previously been calibrated by using a standard initial position and a prescribed motion to correct for mounting uncertainties [10], [1]. However, patients with motor deficits may not be able to achieve these initial positions or perform prescribed movements to produce the suggested joint-to-sensor transformation. Hence, as an alternative, we developed an *in-situ* solution for accurate placement of sensors that is applicable to patients with real-world movement constraints. Specifically, the sensors LA, RA, LF, and RF are placed at their corresponding distal joint segments as shown in Figure 2-4. The carrying angle at the elbow joints and the internal-external rotation at the shoulder joints are displayed in real-time during mounting of the sensors. The sensors are placed correctly when the carrying angle is reflected accurately based on the subject's gender (8º–20º) and internal-external rotations of the LA and RA sensors read zero. This directed real-time mounting strategy can permit correct positioning of sensors without the need to achieve any specific initial position or perform prescribed movements and does not require training in MOCAP.

## 2.2 Exergame and its interfaces

This section presents the main contribution of this chapter, which is also reported in [27]. We have developed an exergame environment using the Unity software application to retrieve the data from the WIS modules and display the same with unique interfaces: (*i*) calibration UI for visualization and assistance with sensor calibration; (*ii*) sensor mounting UI for guided mounting of WIS modules on the human body; (*iii*) patient UI for practicing ROM exercises; (*iv*) playback UI for visualization of patient's performance by clinicians; and (v) instructor UI for creation of customized exercises for patients.

**Calibration user-interface**

The WIS modules consist of tri-axial MARG sensors with each sensor yielding a calibration status ranging from zero (calibration not initialized) to three (all three axes calibrated). An intuitive design with horizontal progress bars is used to represent the calibration status for all the WIS modules. A 3D printed calibration or sensor holder is designed to house the five WIS modules and perform the calibration. The calibration routine for each sensors requires a specific procedure as follows: (*i*) the sensor holder needs to be placed stationary for calibration of the gyroscope; (*ii*) the sensor holder needs to be rotated to ≈45° about each axis for calibration of the accelerometer; and (*iii*) the sensor holder requires random movement in 3D space for calibration of the magnetometer. The users can utilize the visual feedback representing the calibration status of the sensors to perform the calibration routine swiftly. Figure 2-5 shows the calibration UI with indicators for five WIS modules, each with three sensors, with horizontal progress bars that have a discrete resolution of zero (full grey), one/two (partial grey/green), and three (full green).

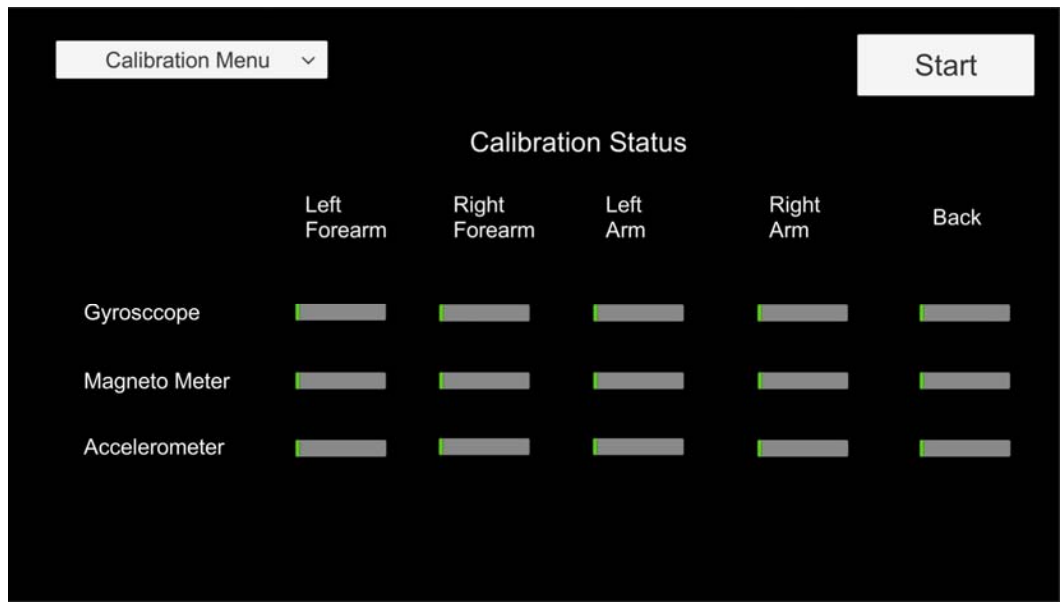

Figure 2-5: Sensor calibration user-interface for visualization of sensor calibration status of five wearable inertial sensor modules each containing three MARG sensors

**Sensor mounting user-interface**

Precise sensor mounting on the human body is critical for accurate measurement of joint ROM. Prior literature has explored the use of standard initial position [3] and determination of joint-to-sensor transformation using specific pre-determined movements prescribed to the user [1]. However, users with movement limitations may not be able to achieve a standard start pose or perform specific actions for anatomical calibration. An *in situ* technique can be used to mount WIS modules to human body for accurate measurement of joint angles [21]. The technique to mount the sensor on the forearm is intuitive due to the anatomical landmark created by the wrist joint on the forearm. However, the sensor placement on the upper arm segment proximal to the elbow requires precise mounting, which is difficult due to skin movements that result in erroneous internal-external rotation angles. To address this challenge, a UI is created for the users to visualize all the JCS joint angles (shoulder: plane, elevation, and internal-external rotation; elbow: flexion-extension, pronation-supination, and carrying angle). Next, the orientation of the UE joints in 3D space is replicated by an animated human model and the UI provides visual cues for rotating the left arm (LA) and right arm (RA) sensors until shoulder internal-external rotation is within $\pm 5°$ for the neutral pose. These visual cues allow the user to adjust the LA and RA sensors to achieve precise alignment with the arm. Figure 2-6 shows the sensor mounting UI with an animated model in neutral pose and the directional cues for the LA and RA sensors.

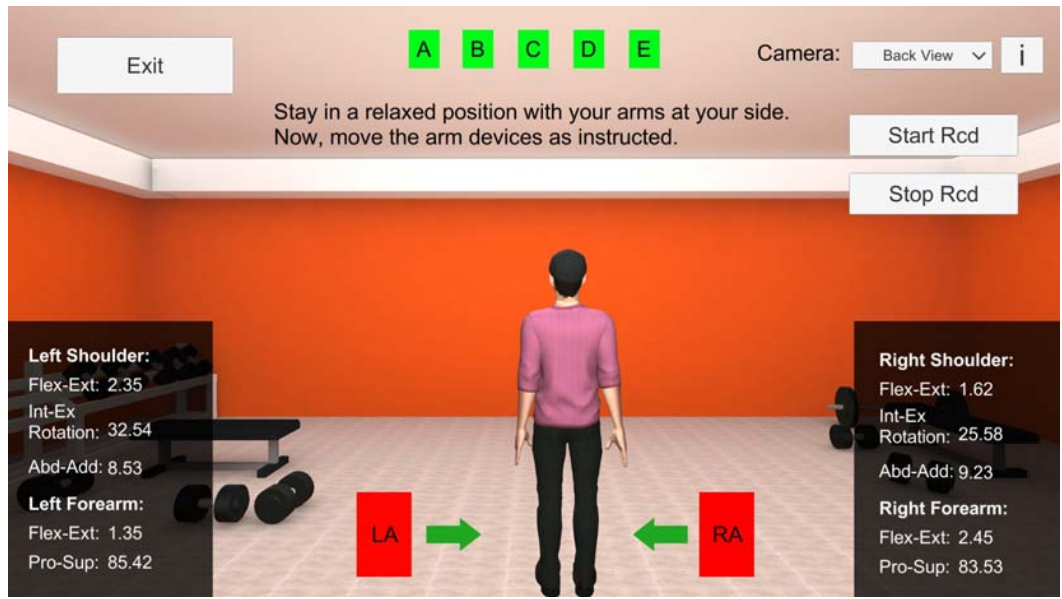

Figure 2-6: Sensor mounting interface showing directional cues for adjusting the sensor mounting on the subject's body

**Patient user-interface**

A game environment emulating a virtual gym is developed for users to practice rehabilitation exercises. The virtual gym includes two human models that can be animated: (*i*) patient and (*ii*) instructor, both in the 3D environment allowing real-time visualization of their movements. The WIS module's absolute quaternions are wirelessly streamed and converted to relative quaternions of the shoulder and forearm movements, which are in-turn utilized to compute the JCS-based joint angles and ROM as in [21]. The UI facilitates a drop-down menu for selecting ROM exercises such as shoulder abduction-adduction, flexion-extension, forearm pronation-supination, etc. The relative quaternions are utilized to animate the patient model to provide real-time visual feedback on the movements being performed. The instructor model demonstrates the ROM exercise selected by the user. During a typical treatment session, the user observes the instructor model performing the selected ROM exercise and examines his/her own movements reflected on the patient model with the data streamed from the WIS modules. Since the movements are reflected on a standard model, the data is de-identified. The interface also provides different viewing angles such as front, back, left, and right views. A screenshot of the developed application showing the instructor and patient user from the back view is presented in Figure 2-7. The data from the patient UI is captured in the JCS framework as quaternions and saved for off-line asynchronous playback and evaluation by the clinicians at their convenience.

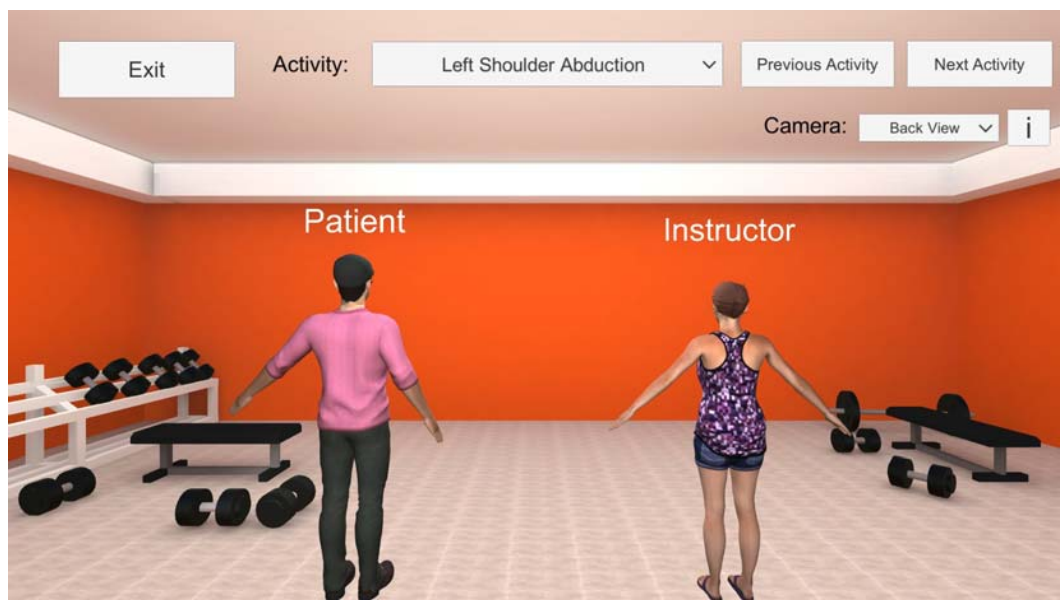

Figure 2-7: Patient user-interface with human models of patient and instructor performing shoulder abduction-adduction movements

**Playback user-interface**

The playback UI facilitates the replay of the recorded ROM activity using the patient UI. The saved quaternion data is unpacked to create an interface similar to a media player with pause and play buttons. Additionally, a seek bar allows the clinician to navigate to specific temporal locations during the exercise for detailed examination of the movement. All the joint angles computed and saved during the exercise are displayed on the left and right panels corresponding to their respective joint angles. An information button "i" allows the user to toggle on/off the display of the joint angles. The camera view can be changed using the dropdown menu on top left (available views: back, front, left, and right). In Figure 2-8, the playback interface shows a patient human model performing an exercise, the joint angles are displayed on each side.

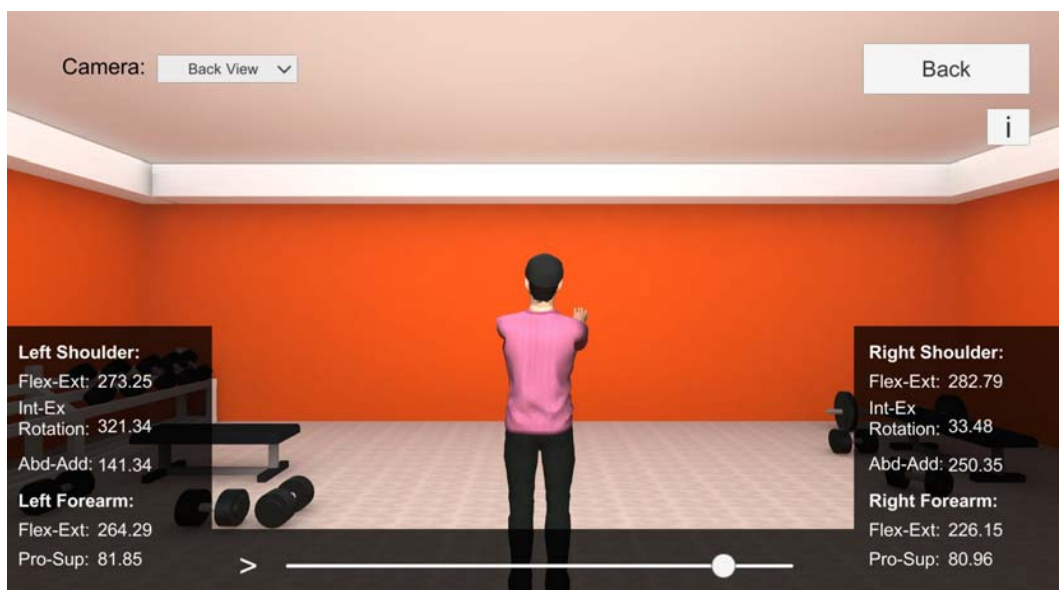

Figure 2-8: Playback user-interface for visualization of patient performance by the therapist

**Instructor user-interface**

An instructor UI is designed for clinicians to develop exercises that are personalized for individual patients based on their therapy needs. This UI consists of a virtual human skeletal model that utilizes the relative quaternions between the sensors to determine joint movements. It allows the clinician to enter the name of the exercise, select key points in the movement, and the time interval between the key points. Each key point saves joint positions of the UE enabling the clinician to create the desired exercises with very little effort. Once completed and saved, the exercise routine consists of the arm passing through the key points, with a set time interval between key points. Spherical linear interpolation, a Unity built-in quaternion interpolation, is performed between the key points for the set time interval to facilitate a smooth movement between all the key points from start to end. A screenshot of the instructor UI with an exoskeleton human model for adding key points is shown in Figure 2-9.

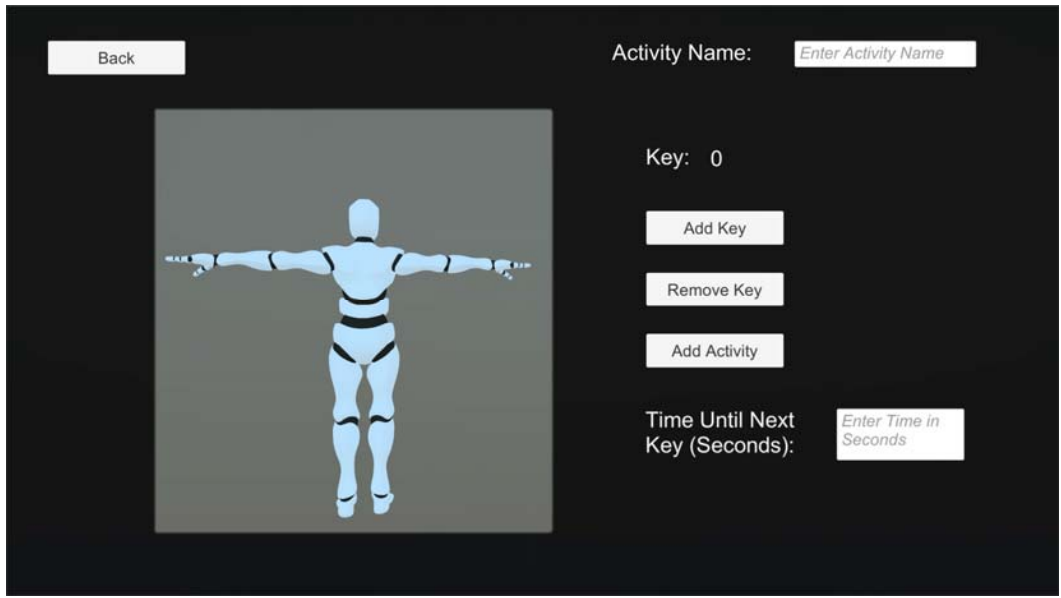

Figure 2-9: Instructor user-interface for creation of ROM exercises by therapists/clinicians

## 2.3 Experimental study

To validate the effectiveness of the developed UIs, two experiments were conducted which document the improvements in (*i*) calibration time and (*ii*) time taken for sensor mounting on the human body as well as (*iii*) examination of user adherence to instructor-programmed exercise routines.

**WIS module calibration time**

In [21], we utilized text output to observe the calibration status. To quantify the effectiveness of the newly developed calibration UI, we compared the time taken for calibration between the prior approach, wherein one observes calibration status from text output communicated via serial port, *versus* the use of the calibration UI. With the UI, calibration time was initialized to zero upon turning-on the devices and the completion time was determined upon successful calibration of all the five WIS modules. The procedure was repeated for five trials and the resulting mean µ and standard deviation σ for the time taken with both approaches are given in Table I.

**WIS module mounting time**

In [21], a MATLAB-based real-time animated plotting interface was created to visualize the joint angles, and based on the resulting plot sensor mounting was adjusted. To evaluate the effectiveness of visual cues for sensor mounting on the human body, we compared the time taken for sensor mounting between the MATLAB-based interface versus the newly created sensor mounting UI. The time taken was recorded from the start of the UI to the successful alignment of the device, i.e., once internal-external rotation reaches within $\pm 5°$. The procedure was repeated for five trials and the resulting µ and σ of the time recorded with both approaches are provided in Table I.

Table 2-1: Comparative sensor calibration and mounting times with exergame and prior interface of [12]

| Task | Time taken in sec. for the task ($\mu \pm \sigma$) | |
|---|---|---|
| | *MATLAB-based UI* | *Exergame UI* |
| Sensor calibration | $41.34 \pm 4.17$ | $32.74 \pm 3.61$ |
| Sensor mounting | $23.95 \pm 2.13$ | $13.23 \pm 1.74$ |

## 2.4 Results

The results in Table I indicate that the exergame interfaces improve the system performance. A media file of various exergame UIs is provided at

[https://youtu.be/SRaNKvxGtFY](https://youtu.be/SRaNKvxGtFY), showing the use of an instructor programmed ROM exercise routine being followed by a user. An experimental trial was conducted to assess the user's adherence to an instructor programmed exercise routine involving shoulder abduction-adduction with maximum ROM of ≈90º for six trials. Using the patient UI, the user performed six repetitions of the exercise. The ROM angle achieved by the user across six repetitions was found to have $\mu \pm \sigma$ of $(97.34 \pm 5.12)°$.

## 2.5 Conclusion

This chapter presents the design and development of exergame interfaces for effective use of the WIS system for ROM assessment. Additionally, the exergame interfaces capture triplanar movements crucial for understanding movement limitations in the JCS framework. The WIS system measurements facilitate data-driven methods for telerehabilitation. We recently developed a grasp rehabilitation device for stroke rehabilitation [28] and it is being used for a clinical trial involving patients with multiple sclerosis in a telerehabilitative setting [29]. Similar clinical trials can be performed using the WIS system for commercialization and broader adoption. Unity's capability to deploy applications on smartphones will be leveraged for developing these exergames as smartphone and tablet applications for use by patients and clinicians. Such systems will improve the connectivity between clinicians and patients to facilitate precision rehabilitation.

# Chapter 3 Fork Rehabilitation

## 3.1 Introduction

The ability to control utensils for the purpose of eating is one of the most difficult ADL to perform after a stroke [30], [31]. Furthermore, patients maintain neuroplasticity by engaging in repetitive rehabilitation focused on a specific task [32]. However, limited number of skilled therapists, expensive treatment [33], and lack of patient motivation towards tedious exercises [34] account for lower recovery rates. Current research in robot-assisted rehabilitation [35] and its integration with telerehabilitation and video games have shown a great promise in improving patient's motivation and performance [36], [37]. Commercially available game consoles such as Sony PlayStation, Nintendo Wii and Microsoft Kinect have been explored for enhancing patient's rehabilitation experience [38].

We have integrated a game environment with low-cost mechatronics-based eating utensils (see Figure 3-1) to develop a system called RehabFork. RehabFork focuses entertaining the patients while performing repetitive functional sub-tasks of eating such as (*i*) grasping the utensils; (*ii*) lifting the utensils; and (*iii*) holding and cutting the food using fork and knife. These activities are a part of twenty vital ADLs under Functional Task Battery [39] that utilize the upper limb(s).

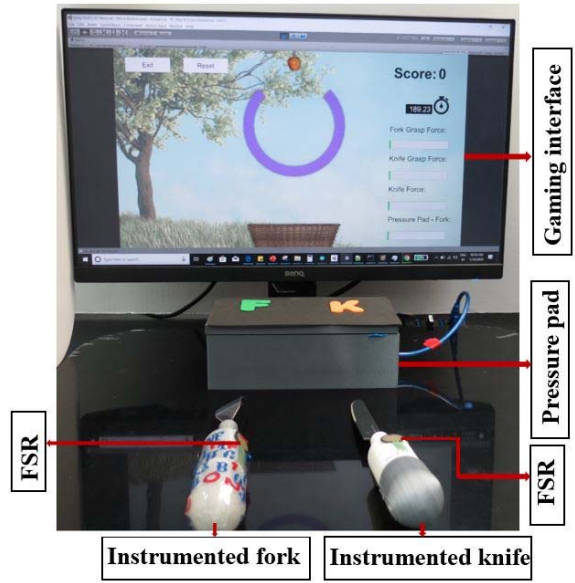

Figure 3-1: Prototype of upper-extremity rehabilitation system consisting of an instrumented fork and knife, a 3D printed pressure pad, and an interactive gaming interface

## 3.2 RehabFork system

This section is adapted from [40] and included for completeness. The RehabFork system includes an exergame integrated with an instrumentation fork, an instrumented knife, and 3D printed pressure pad. Figure 3-1 shows the RehabFork system. Every sub-task of eating is enhanced by the exergame that provides a real-time visual feedback to post-stroke patients (henceforth called users) utilizing the sensors embedded on the utensils. This feedback includes information such as wrist rotation angle and forces applied during grasping, poking, and cutting (see Figure 3-2).

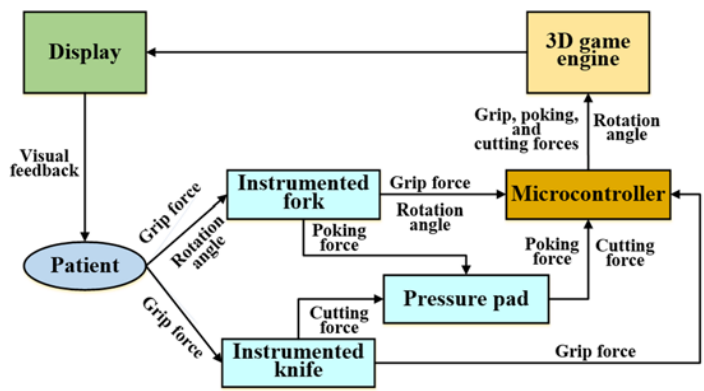

Figure 3-2: Framework of upper-extremity rehabilitation system

**Instrumented Fork**

Two Interlink 402 force sensitive resistor (FSR) sensors and one BNO055 Inertial Measurement Unit (IMU) are attached to the fork handle cap. The IMU, housed inside the fork handle cap, is used to obtain the rotation measurements of the user's forearm when the user manipulates the fork. A Bluno Beetle, an Arduino-based Bluetooth Low Energy (BLE)-enabled microcontroller, housed inside the fork handle cap, is used to acquire and process the FSR and IMU data of the fork and to wirelessly transmit these data to the Unity game engine.

**Instrumented Knife**

To obtain the knife grasp force value, two Interlink 402 FSR sensors are attached to the handle cap of the knife. Another Bluno Beetle microcontroller, housed inside the knife handle cap, is used to acquire and process the FSR data of the knife and to wirelessly transmit these measurements to an Arduino Nano microcontroller of the pressure pad for further transmission to the Unity game engine.

**Pressure pad**

The pressure pad is a 3D printed box that consists of separate regions designed for fork and knife actions. Two square Interlink 406 FSR sensors are attached to the base of the fork and knife regions. The fork region is used to obtain the force that the user applies to this area of the pressure pad as they hold it down using the fork to simulate the action of holding down food. The knife region similarly consists of another FSR that obtains the force the user applies to that area of the pressure pad to simulate the action of cutting food. The pressure pad is a wired system consisting of an Arduino Nano microcontroller that processes the FSR data for the poking and cutting forces and transmits these measurements, concatenated with the measurements received from the instrumented knife, to the Unity game engine through a USB connection.

### 3.3 Exergame and interfaces

In this section, we present the main contribution of this chapter. Specifically, we provide the details of the designed gaming interface that maintains user motivation and engagement as they train their upper limb to perform multiple repetitions of certain sub-tasks of eating. The target population includes patients with varying severity of upper-limb disabilities and may include elderly patients at home. Keeping this in mind, the gaming environment developed using the Unity game engine is deployable on a PC or laptop and it does not require high system requirements. As the user manipulates the instrumented fork and knife along with the pressure pad, the interactive gaming interface displays a combination of game objects that respond to the user's actions. The developed gaming interface consists of three progressive levels which are described below.

**Level 1 and Level 2**

Levels 1 and 2 of the game (see Figure 3-3) utilize the instrumented fork with the pressure pad. The instrumented knife is not utilized during game play in these levels. The force bar on the screen displays the amount of force (on a linear scale) that the user needs to apply to grasp the fork. This required force corresponds to force parameter values set during the calibration step.

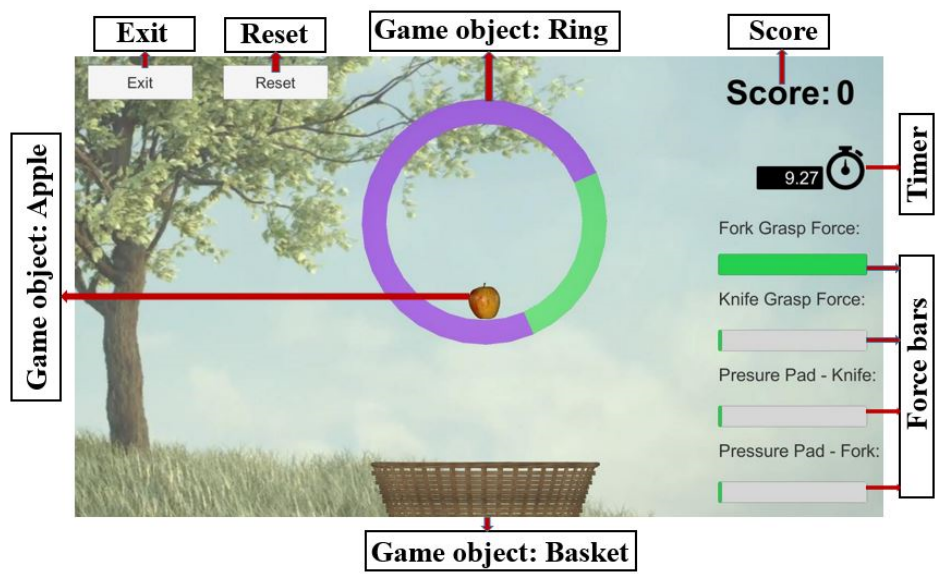

Figure 3-3: Levels 1 and 2 gaming interfaces

The game scene in Levels 1 and 2 contain three objects–a Ring, an Apple, and a Basket. Once the user grasps the fork with sufficient force, the top of the 'Ring' game object opens up and the 'Apple' game object enters the ring. Once the user lifts the fork, rotates their forearm to at least 90°, and points the fork towards the pressure pad, the ring closes and rotates such that the opening of the ring points towards the 'Basket' game object placed on the ground. The rotation of the ring is driven using the IMU sensor data for the angular rotation of the forearm. As the user 'pokes' down on the pressure pad with the fork, the ring opens up and the apple falls into the basket. This completes Level 1 of game and the user is shown the time taken to complete the activity and score. To progress to Level 2, the user has to successfully complete the Level 1 activity three times. Level 2 follows the same mechanism as Level 1. In Level 1, the minimum force required to grasp the instrumented fork is calculated as 50% of the average calibration force of the unaffected hand. In Level 2, this is increased to 75% of the average calibration force of the unaffected hand. The user also needs to orient the fork with respect to the fork region of the pressure pad more accurately (rotate the forearm at least 135°) to complete Level 2 successfully. To progress to Level 3, the user has to successfully complete the Level 2 activity six times.

**Level 3**

In Level 3 (see Figure 3-4), both the instrumented fork and knife pathways are utilized via the pressure pad and a fourth game object, an Apple box, is included. The first stage of Level 3 resembles the previous levels, i.e., the user grasps the fork, lifts it, rotates it, and positions it with respect to the pressure pad. In response, the apple game object on the screen moves in the ring, the ring rotates, the ring opens up, and the apple falls into the 'Apple box' game object floating above the basket. In the second stage of Level 3, the user grasps the knife by applying at least the minimum force as displayed on the corresponding force bar on the screen. The bars in Level 3 display 100% of the calibrated force value of the unaffected hand. Once the user has successfully grasped the knife, they lift and place it on the designated knife region of the pressure pad and simulate the cutting motion. This force value from the action is transmitted to the game engine by the microcontroller and it triggers the apple to slide down into the basket. At this stage, the user needs to perform the task nine times to complete the activity. A timer records and displays the task completion time, serving as a metric of user performance in addition to other game metrics, namely, *(i)* scores and *(ii)* progression in levels. Finally, all the user information and their performance data (including scores, levels of game completed, time taken to complete each activity, grasp quality, and poking/cutting force values, etc.) are stored in a database and the therapist can access this information to monitor the user's progress.

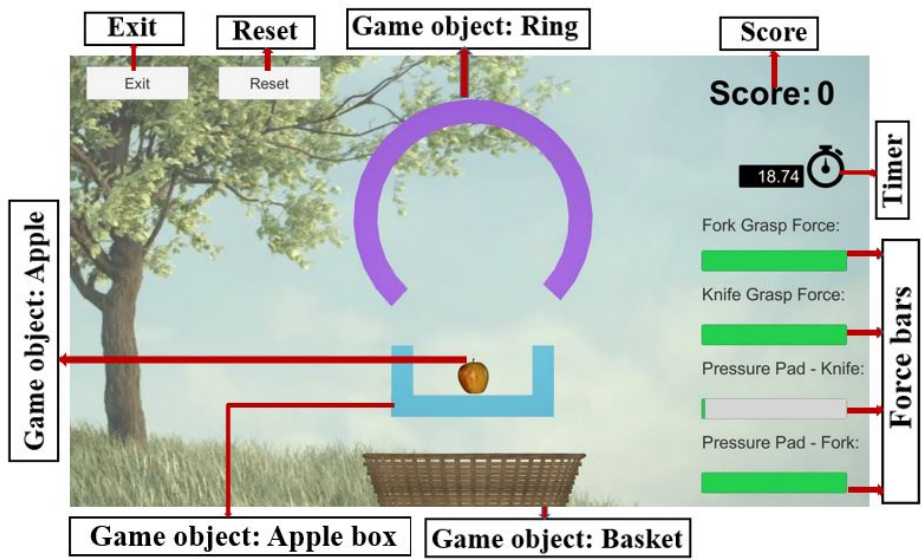

Figure 3-4: Level 3 gaming interface

## 3.4 Conclusion

An exergame for task-specific upper limb rehabilitation system, RehabFork, is designed and developed. The exergame include different game levels that entertain users, provide visual feedback and gauge user performance. The selection of these levels can be done by a clinician based on patient's performance. The RehabFork system enables home-based rehabilitation for post-stroke patients.

# Chapter 4 Grasp rehabilitation

## 4.1 Introduction

Grasping and manipulation are integral for performing activities of daily living. During a simple grasp and lift task, the object attributes such as shape, size, and texture affect the fingertip forces exerted by the individual [41]–[44]. Healthy individuals apply controlled fingertip grasping forces preventing slippage and over squeezing of the object based on the sensory feedback. A neurological event, such as stroke, spine injury, or nerve damage, can affect one's motor skills also leading to loss of the ability to control the fingertip forces [44]–[47]. Several commercially available devices including MusicGlove [48], Hand tutor [48], Hand mentor [48], and Raphael smart glove [48] are used for hand rehabilitation but none of them provide feedback of fingertip forces. To solve this, we developed an exergame interface for a portable, mechatronic-based grasp rehabilitation device that provides a visual feedback to the user input through a videogame.

## 4.2 Mechatronics grasp rehabilitation device

The mechatronics grasp rehabilitation device uses off-the-shelf sensors such as load cells for acquiring the grasp and lifting forces [28]. Following [44], the force sensing was restricted to within 0-10N and it was found that load cells fitted with strain gauges provided a high-resolution force sensing in this range. The grasp rehabilitation device consisted of two circular grasping surfaces. Each of these surfaces was attached to a load cell which enabled measurement of grip forces. Various textured 3D printed caps were utilized to augment surfaces with different textures. The device's weight can also be modified using a space included in the device where a 3D printed drawer can be inserted with desired weight in it. Furthermore, the device consists of a load cell on the bottom to measure the lifting forces when the device is lifted. The grasp rehabilitation device with various textures and weighted drawers are shown in Figure 4-1. All the weighed drawers have uniform shape and size to prevent the user from knowing the actual weight. These drawers are used to simulate objects with different weights.

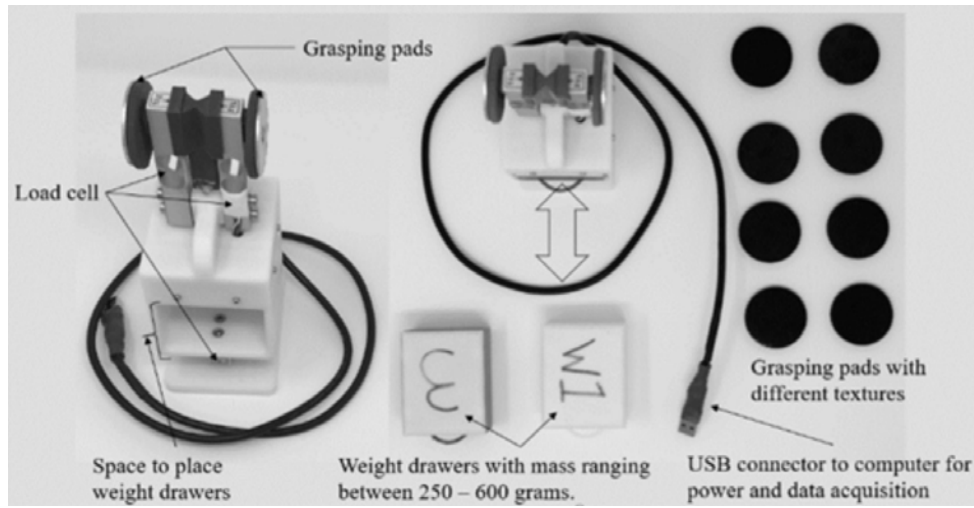

Figure 4-1: The grasp rehabilitation device with various textures and weighted drawers

## 4.3 Exergame and its interfaces

We have developed a Unity game engine based exergame, which utilizes the grasping and lifting forces provided by the grasp rehabilitation device. This game consists of an astronaut as player's avatar who runs through space, hopping from one space rock (platform) to another. The jumping action in this game is proportional to the lifting force applied and the grasping force causes the astronaut to expand and eventually explode (only in Level 3) when the grasping force exceeds a preset limit (when this happens, the game ends in a failure). An indicator of grasp force is provided at the top of the screen to offer a visual feedback for the amount of grasp force being applied. The objective of the game is to perform the grasp task properly that facilitates the astronaut to move from one rock to another until the player reaches his rocket (the rocket only spawns on the space rocks). Furthermore, the game has two layers of platforms. The player must stay on the space rocks (top platform) to collect stars and finish the game. If the player fails to make the jump, s/he falls to the ground (bottom platform) where no stars are available. The player must move back to the space rocks by jumping into the warp holes to be able to finish the game successfully. The warp holes appear on the ground and have the sole purpose of transporting the player to the space rocks. Additionally, there are holes that appear along the ground, falling through them will end the game in a failure. This game has three levels of difficulties that can be changed based on player's performance.

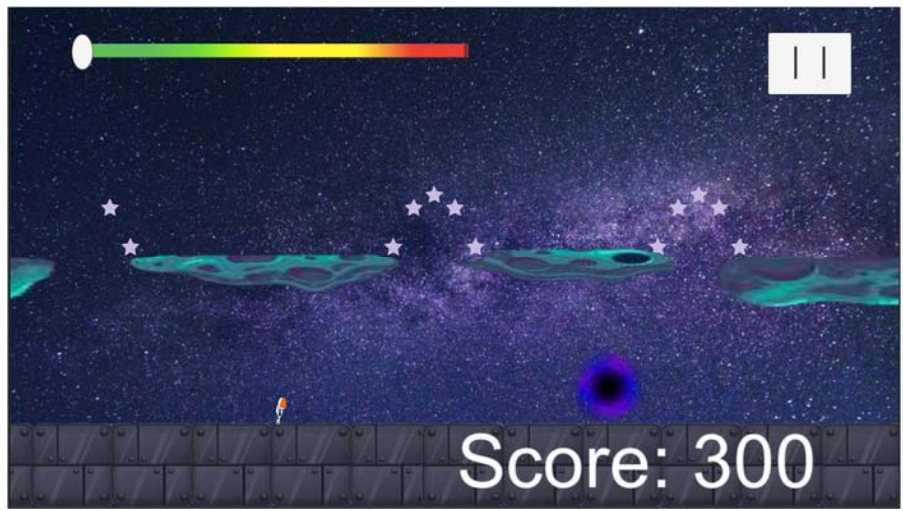

Figure 4-2: Level 1 game interface

**Level 1**

Level 1 is the easiest of all the levels. There are no holes along the ground and the astronaut does not explode when grasp force exceeds the limit. This ensures that the player cannot lose the game due to external factors and has to perform appropriate lift task to win the level. This level aims to help patients get accustomed to the game along with lifting task. Here, the lift task takes the most priority. Level 1 interface is shown in Figure 4-2.

**Level 2**

Level 2 adds the first challenge by adding holes along the ground. This adds the first external element that can cause the player to fail. Additionally, the astronaut's running speed increases by 50%. The distance between the rocks is also increased to make it slightly difficult to jump than level 1. The astronaut still does not explode when grasp force exceeds the limit. So, the lift task still takes the most priority. Level 2 interface is shown in Figure 4-3.

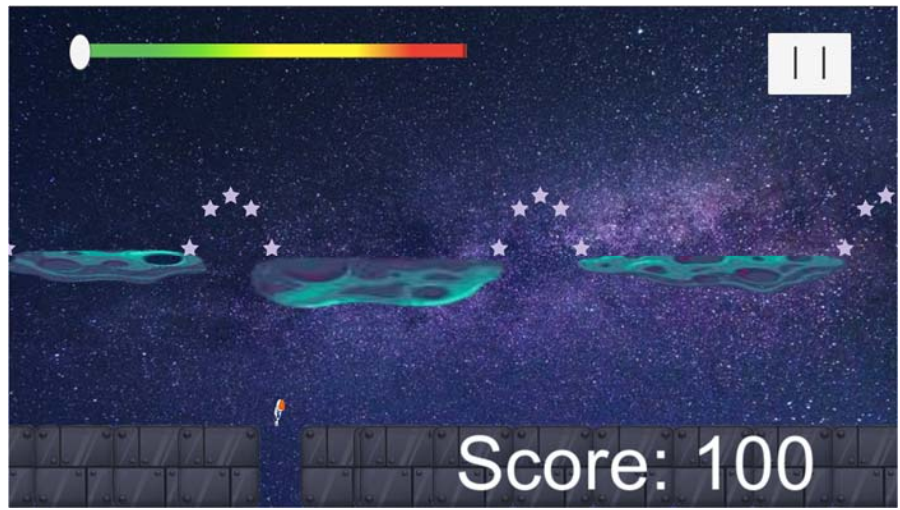

Figure 4-3: Level 2 game interface

**Level 3**

Level 3 is the most challenging, in addition to having holes along the ground, the astronaut now explodes when grasp force exceeds the limit. So, the player must regulate both grasp and lift forces accordingly. Moreover, the astronaut's running speed increases by 100% when compared to Level 1 and the distance between the rocks is further increased. Level 3 interface is shown in Figure 4-4.

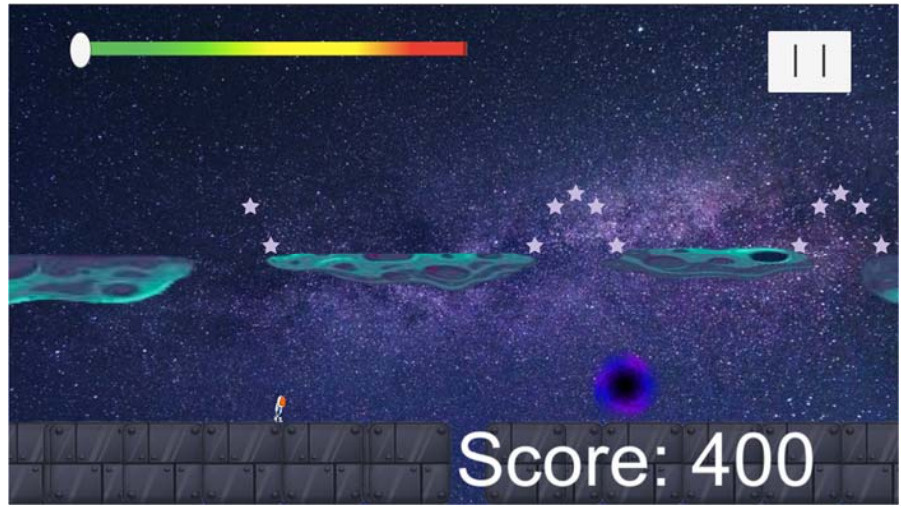

Figure 4-4: Level 3 game interface

## 4.4 Conclusion

The design and development of an exergame environment for low-cost grasp rehabilitation device was presented. Furthermore, the grasping and lifting forces of the device have been translated to create an entertaining video game. This game intends to provide an enhanced feedback to the patient through various game mechanics. Moreover, the game allows the clinician to gauge the patient's performance and adapt the therapy accordingly.

# Chapter 5 Contributions and Future Work

Chapter 2 presented the design of various user-interfaces for an exergames that provides an intuitive and efficient interaction with the WIS system. The effectiveness of the developed interfaces is assessed for sensor calibration and mounting. The results indicate that the designed user interfaces improve the usability of the WIS system. The work presented in Chapter 2 was developed in collaboration with A. RajKumar, PhD student, and F. Vulpi, M.S. student, in Mechatronics, Controls, and Robotics Laboratory. Materials critical to the development of this thesis are included in Chapter 2, with some unavoidable overlap from [49], [50].

Chapter 3 presented the design of gaming environments for task-specific rehabilitation of holding the food, cutting, and eating using the RehabFork System. The potential of the system was assessed through a user-study and questionnaires. The work presented in Chapter 3 was developed in collaboration with V.J. Krishnan, PhD student, and Sahil Kumar, M.S. student, in Mechatronics, Controls, and Robotics Laboratory. Materials critical to the development of this thesis are included in Chapter 3, with some unavoidable overlap from [51].

Chapter 4 presented the design of an exergame performing lifting and grasping tasks using grasp rehabilitation device. The works presented in Chapter 4 was developed in collaboration with A. RajKumar, PhD student, in Mechatronics, Controls, and Robotics Laboratory. Materials critical to the development of this thesis are included in Chapter 4, with some unavoidable overlap from [50].

This thesis has illustrated the design and development of exergames for low-cost mechatronic devices for home-based rehabilitation. In future research, for Chapter 2, the avatars can be improved, and more default exercises may be provided. Additional options should be provided to include external objects such as sticks, balls, and bands. More devices should be added to capture the entire body motion. For Chapter 3, adaptive scoring system can be implemented to scale the difficulty according to the performance. Sensor fusion using filters, such as extended Kalman filter or particle filter, can be used to improve the gameplay by estimating the y-axis displacement of the fork. For Chapters 2, 3, and 4, some general improvements can be directed towards: (*i*) conducting user studies with disabled individuals/stroke patient after getting IRB approval; (*ii*) utilizing the information and feedback from the patients to mitigate flaws and improve the game and device; (*iii*) exploring Unity's ability to build for multiple platforms and making the system portable for patients and clinicians by deploying the game on smartphones; (*iv*) replacing Bluetooth with WiFi connectivity for the benefits of increased number of connections, range, and speed; and (*v*) using robust and higher quality sensors for more reliability.